\documentclass[review,authoryear]{elsarticle}
\journal{Transportation Research Part C}
\usepackage{amsmath}
\usepackage[mathlines]{lineno}
\usepackage{amsfonts}
\usepackage{amssymb}
\usepackage{graphicx}
\usepackage{multirow}
\usepackage[table,xcdraw]{xcolor}
\usepackage{float}
\usepackage{verbatim}
\usepackage{hyperref} 
\usepackage{fancyhdr} 

\usepackage{makecell} 
\usepackage{hhline} 

\immediate\write18{texcount -tex -sum  \jobname.tex > \jobname.wordcount.tex}

\date{} 

\begin{document}

\begin{frontmatter}
\title{Traffic congestion and travel time prediction\\
based on historical congestion maps and identification of consensual days}

\author[LICIT,IRT]{Nicolas Chiabaut \corref{cor}}
\author[SPIE,IRT]{Rémi Faitout}

\cortext[cor]{Corresponding author.}

\address[LICIT]{Univ. Lyon, ENTPE, IFSTTAR, LICIT, F-69518, Lyon, France}
\address[SPIE]{SPIE CityNetworks, F-69320, Feyzin, France}
\address[IRT]{IRT System-X, Lyon, France}

\begin{abstract}
In this paper, a new practice-ready method for the real-time estimation of traffic conditions and travel times on highways is introduced. 
First, after a principal component analysis, observation days of a historical dataset are clustered. 
Two different methods are compared: a Gaussian Mixture Model and a k-means algorithm.
The clustering results reveal that congestion maps of days of the same group have substantial similarity in their traffic conditions and dynamic.
Such a map is a binary visualization of the congestion propagation on the freeway, giving more importance to the traffic dynamics.
Second, a consensus day is identified in each cluster as the most representative day of the community according to the congestion maps. 
Third, this information obtained from the historical data is used to predict traffic congestion propagation and travel times.
Thus, the first measurements of a new day are used to determine which consensual day is the closest to this new day. 
The past observations recorded for that consensual day are then used to predict future traffic conditions and travel times. 
This method is tested using ten months of data collected on a French freeway and shows very encouraging results.
\end{abstract}


\end{frontmatter}


\section{Introduction}

Prediction of traffic states and travel times evolution is a key component of any traffic monitoring system and decision support system.
Their accurate estimation is critical for freeway managers, especially when the network becomes congested. 
This problem has been extensively investigated in the transportation literature using model-based, simulation-based and data-driven approaches \citep{VLAHOGIANNI2005211,mori_review_2015,Wang2018}. 
For short-time prediction, model-based and simulation-based approaches use traffic flow models in conjunction with data assimilation techniques such as recursive Bayesian estimators to predict the traffic states and the resulting travel times \citep{vlahogianni_short-term_2014,mori_review_2015,wang_renaissance_2006,kumar_bus_2017}.
Most data-driven approaches use general purpose parameterized mathematical model such as linear regression \citep{rice_simple_2004}, Kalman filtering \citep{van_lint_online_2008,Nanthawichit2003}, particle filters \citep{Wang2007} support vector regression \citep{huang_deep_2014}, random forest, Bayesian network \citep{LI2019}, artificial neural networks \citep{adeli2001,van_lint_online_2008,VLAHOGIANNI2005211,XU2020,LI2017} and many other techniques to capture and learn from data the correlations between traffic variables (speed, travel-time) over space and time \citep{COIFMAN2002351,Polson2017,MA2020}.
As pointed out by \cite{yildirimoglu_experienced_2013}, who wrote a complete and useful state of the art of the estimation methods, these approaches suffer from various limitations.
To quote only a few, a common limitation is the spatio-temporal correlations that are mainly artificially selected \citep{XU2020}.
Beside, some of the existing methods resort to experienced travel times, i.e. travel times calculated by traveling a trajectory through the velocity field. 
However, this information is rarely available in real time because experienced travel time is usually greater than the prediction horizon.

\begin{figure}[H]
    \centering
        \includegraphics[width=\textwidth]{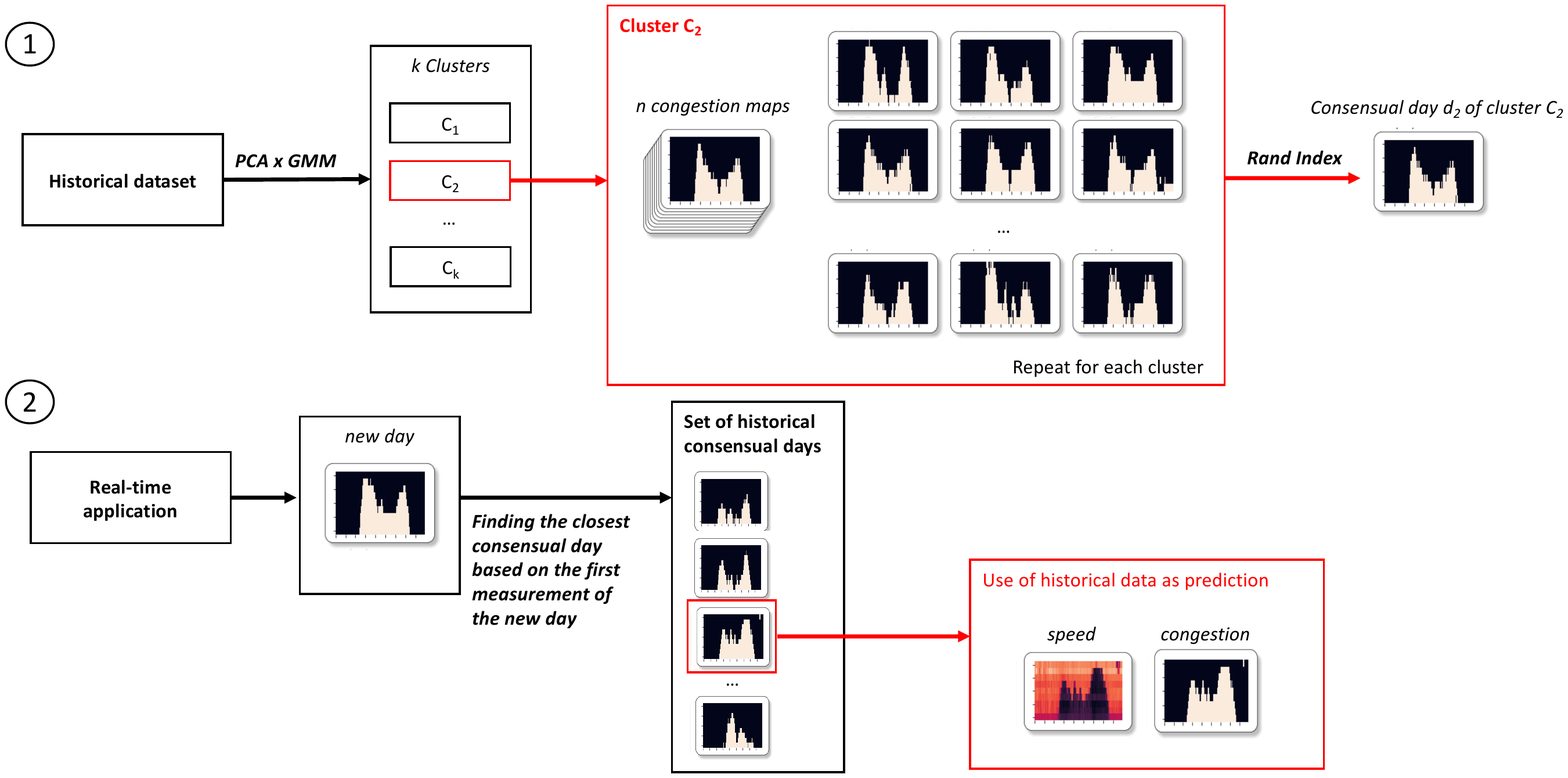}
        \caption{Graphical representation of the proposed method}
        \label{fig:fig0}
\end{figure}

Consequently, the purpose of this paper is to the evolution of congestion, therefore, travel times with a simple, fully explainable, and practice-ready method.
The proposed method uses both historical and real time traffic information to calculate short-term congestion and travel time evolution forecast. 
To this end, we used the concept of \textit{congestion map} to consider queue propagation rather than traffic states variables evolution, such as density or speed.
Figure \ref{fig:fig0} presents the mechanism of the global algorithm of the proposed method.

The fist step (\textcircled{\tiny{1}}) consists in partitioning historical information into $k$ clusters $C_k$ presenting similar characteristics based on the traffic patterns observed in the highway.
As in \cite{yildirimoglu_experienced_2013}, \cite{SHIVANAGENDRA2003285} and many other papers, we first resort to a Principal Component Analysis to reduce the number of variables.
Then, a Gaussian Mixture Model and a k-means algorithm are used to gather the historical data.
Note that both approaches are used to evaluate the sensitivity of the estimation method to the clustering process.
Then, the proposed method differs from the one of \cite{yildirimoglu_experienced_2013}: rather than considering the global behavior of the clusters, we try to identify which day $d_k$ within the cluster $C_k$ is the most representative of the group.  
This so-called \textit{consensual day} is determined based on the congestion maps of the clustered days through a method derived from consensual learning technique \citep{FILKOV2004}.

Once a set of consensual days $D_k$ has been established, the second step of the method (\textcircled{\tiny{2}}) is devoted to real-time application.
Based on the first minutes / hours of a new observation day, measurements are processed in real-time to identify in this set which is the closest consensual day in $D_k$. 
The recorded congestion map and the observed speed of this closest consensual day are then used to predict the congestion and travel times evolution of the next minutes / hours of the new observation day.
The main benefit of using past measurements to compute future traffic conditions is to ensure realistic values that are fully consistent with traffic dynamics.

The remainder of the paper is organized as follows: Section 2 presents the case study and the dataset used in the paper, Section 3 introduces the prediction methods, Section 4 is devoted to the clustering process and Section 5 to the analysis of the results, while Section 6 includes a conclusion.

\section{Case study and dataset}

\begin{figure}[H]
    \centering
        \includegraphics[width=\textwidth]{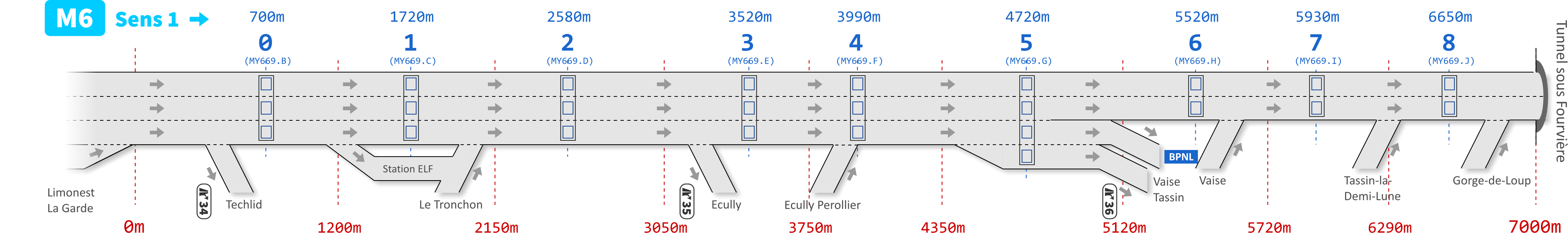}
        \caption{Sketch of the studied site}
        \label{fig:fig1}
\end{figure}

In this paper, we focus on the $M6$ highway near Lyon, France. 
Figure \ref{fig:fig1} depicts a sketch of the site.
It is important to notice that this highway is used to access Lyon's city center through a tunnel, which is a recurrent, active bottleneck.
Moreover, this highway is one of the most important in France and favored by holidaymakers because it links Paris to the south of France (French Riviera) and the Alps. 
This highway is thus also called \textit{Motorway of the Sun}.
Consequently, major traffic jams are always observed during holidays.

Regularly staggered loop detectors can be found on this highway section. 
These detectors provide average flow, speed and occupancy rate per lane every $1$ minute. 
In this study, we mainly focus on data from $9$ detectors of a $7$ km long section, see Figure \ref{fig:fig1}.
Accordingly, $9$ sections of length $\Delta l$ roughly centered on the detectors have been defined. 
The maximal authorized speed varies from $90$ km/h to $70$ km/h within this section.
In the remainder of the paper, detectors are labeled by increasing positions from $0$ to $8$.
We consider that traffic conditions between detectors can be interpolated by using observations of the closest detector.
All data from January $2018$ to October $2018$ is available (except May). 
We partitioned the data by day from $6$:$00$ to $22$:$00$ ($960$ observations per day).
Finally, data has been roughly cleaned up to remove unrealistic values or problem of acquisition.

\section{Methodology}

\subsection{Congestion maps}
\begin{figure}[H]
    \centering
        (a)
        \includegraphics[width=\textwidth]{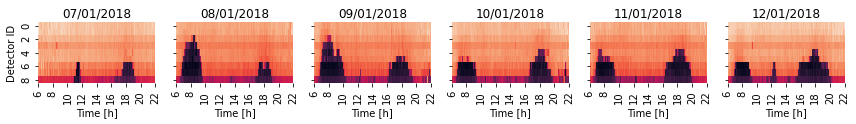}
        (b)
        \includegraphics[width=\textwidth]{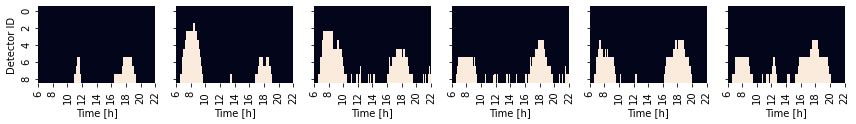}
        \caption{(a) Speed maps for $6$ different days of the historical dataset (traffic flows from 0 to 8), and (b) associated binary congestion maps $M_d$}
        \label{fig:fig2}
\end{figure}

To mainly focus on traffic dynamic rather than speed evolution, we used the concept of \textit{congestion map}.
Indeed, recorded values of speed can vary because of many local phenomenons (such as variations in driving behaviors, noises in measurements, etc.) but without correlation with the macroscopic dynamics of flow that rule traffic conditions.
One potential method, but far from perfect, to reduce this bias is to average the observations.
Here, we decide to use a more drastic method to only focus on two possible traffic states: free-flow and congestion.
For each loop detector $l \in [0,8]$, we consider therefore a variable $x_l$ that, at time $t$, is equal to $1$ if the observed speed $v_l(t)$ is lower than a congested speed threshold $v_{cong}$ (fixed here at $40$ km/h) and equal to $0$ otherwise.
It makes it possible to compute map $M_d$ of day $d$ as Boolean matrix of size "number of detectors" x "number of observations per day" composed of elements $x_l(t)$.
Figure \ref{fig:fig2}a shows the classical speed maps for $6$ randomly selected days of the case study. 
Note that the darker the color the lower the speed and that the traffic flows from the top to the bottom of the graphs.
Different congestion patterns clearly appear with different length propagation and duration. 
These observation are easier when focusing on the associated congestion maps $M_d$ in Figure \ref{fig:fig2}b. 
Note that black color stands for $x_l(t) = 0$ (free-flow) and white color for $x_l(t) = 1$ (congestion). 

\subsection{Clustering historical data}

The regularity of traffic events makes very useful the information that we can obtain from historical data and what can been learned from past situations.
Consequently, it is worth appealing to classify the different observation days.
Because the size of the initial dataset is very large ($8640$ variables for a single day), the first step, to speed up the clustering methods and obtain accurate results, is to perform a Principal Component Analysis (PCA) to reduce the dimensions of the observations \citep{SHIVANAGENDRA2003285}. 
Notice that we consider here speed ${v_l}(t)$ for $l \in [0,8]$ and $t \in [0, 960]$ as the main vector of the PCA.
Since PCA is a usual method, we do not study here the results in details.

Then, the purpose of the second step of the method is to cluster similar days of the historical dataset.
To this end, we use two classical clustering algorithms: (i) the k-means with an Euclidean distance between observations to gather and (ii) a Gaussian Mixture Model (GMM) as proposed in \cite{yildirimoglu_experienced_2013}.

(i) The k-means algorithm is one of the most popular unsupervised learning methods, that aims to gather the data into groups of equal variance that minimize the inertia, i.e., within-cluster sum-of-squares.
With a fast computing time, results are easy to interpret, but k-means implicitly assumes that all clusters are spherical.
This shape could introduce a strong bias, especially for observations of highly non-linear phenomenons.

(ii) In the GMM method, the underlying idea is to consider that similar days constituting the different clusters follow normal distributions.
Consequently, the set of clusters, i.e., the partition, is ruled by a GMM.
This clustering method gives importance to the distribution of the data points and not only the distance between them.
Consequently, GMM is well adapted to our case because it provides clusters that may have different sizes and correlation within them.
Note that a full covariance matrix is assumed in GMM.



\subsection{Identifying consensual days}

Now, the idea is to determine which day of a cluster is the most representative of the group.
As already explained, the motivation is to use the representative days as the prediction for the coming days, see Figure \ref{fig:fig0}.
To this end, we have adapted the consensual learning method used in \cite{lopez_revealing_2017} to our specific case.
Thus, the representative day of a cluster is identified according to a distance based on the Rand index, i.e., the accuracy between congestion maps.
The Rand index between two maps $M_d$ and $M_p$ of days $d$ and $p$ is defined as the number of concomitant results among the total number of observations.
Here, we consider two observations as concomitant when, at a given detector $l$ and time $t$, both observations for day $d$ and $p$ are in the same state (free-flow / free-flow or congested / congested), i.e. $M_d(i,j) = M_p(i,j)$.
The distance that we used can be formulated as:

\begin{equation}
    R_M(d,p) = \frac{a}{a+b+c}
\end{equation}
where:
\begin{itemize}
    \item $a$ is the size of $\{x_l(t), M_d(l,t)-M_p(l,t) = 0 \}$ (free-flow / free-flow or congested / congested);
    \item $b = \vert \{x_l(t), M_d(l,t)-M_p(l,t) = 1\} \vert $ (free-flow / congested);
    \item $c = \vert \{x_l(t), M_d(l,t)-M_p(l,t) = -1\}\vert$ (congested / free-flow);
\end{itemize}

Then, we define the consensual day $d_k$ of a given cluster $C_k$ as the one that maximizes the sum of the Rand indices within a cluster:

\begin{equation}
    d_k = \text{arg}\,\max_{d \in C_k} \{\sum_{p \in C_k} R_M(d,p)\}
\end{equation}

Consequently, the set of consensual days $D_k$ can be determined for the whole historical dataset.

\subsection{Travel time prediction}

We are now going to take advantage of the consensual days to predict both congestion and travel times evolution in real time.
Consider a new day of observation $p$.
The time interval is discretized into periods of $\delta t$ minutes.
At time $t$, we try to determine which consensual day $d_k \in D_k$ is the closest to this new day $p$.
To this end, we only consider the last $\Delta t$ observations and build a partial congestion map $m_p(t-\Delta t,t)$ composed of $x_l(t)$ for $t \in [t-\Delta t,t]$ and $l \in [0,8]$.
This map is compared to the partial maps extracted from the congestion maps of the consensual days $m_{d_k}(t-\Delta t,t) = M_{d_k}(t'), \: \forall t' / t' \in [t-\Delta t,t] $ and $d_k \in D_k$.
The consensual day $d^*_p(t)$ that has the maximal Rand index, based on the partial maps, with the observation $p$ is selected: 

\begin{equation}
    d^*_p(t) = \text{arg}\,\max_{d \in d_k} \{R_m(p,d_k)\}    
\end{equation}

Now that $d^*_p(t)$ have been identified, we used the historical data $x_l(t)$ and $v_l(t)$ of $d^*_p$ to predict variables of day $p$ for the next time step $\delta t$.
This process is iterated at every $\delta t$, and congestion and speed maps can be built by gathering the data.
Notice that the optimal consensual day $d^*_p$ can (and surely will) change with time $t$.
Duration $\delta t$ (prediction horizon) and $\Delta t$ (learning period) belong to the parameters of the proposed method.
Thus, the prediction of the congestion map and the travel times (respectively) of day $p$ are (respectively):
\begin{equation}
    M^*_p(l,t) = M_{d^*_p(t)}(l,t), \: \forall l \in [0,8]     
\end{equation}
and:
\begin{equation}
    v^*(l,t) = v_{d^*_p(t)}(l,t).    
\end{equation}
where $M_{d^*_p(t)}(l,t)$ is the element of the observed congestion map of the consensual day $d^*_p(t)$ at time $t$ for detector $l$, and $v_{d^*_p(t)}(l,t)$ is the observed speed at time $t$ on the congestion day $d^*_p(t)$ for detector $l$.

The travel time $\tau$ at time $t$ is calculated as:
\begin{equation}
    \tau(t) = \sum_{l=0}^{9} \frac{\Delta x_{l}}{v_l(t)}
\end{equation}
where $\Delta x_{l}$ is the length of section $l$, see Figure \ref{fig:fig1}.
It is important to notice that it corresponds to the definition of the instantaneous travel times: the travel time at time $t$ is calculated based on the speed at the different detector locations at time $t$ \citep{YEON2008325}.
It may introduce some bias compared to the experience travel times, but, in our case, this bias is very limited because of the relatively short length of the case study (less than $10$ km).

In order to evaluate the global prediction method proposed in the paper, a cross-validation procedure can be used. 
A simple holdout method is considered by randomly selecting $75 \%$ of the initial data as the learning set.
This training set is clustered into $K$ groups, and the associated $K$ consensual days are determined.
Then, travel times are predicted for the remaining $25 \%$ and compared with the observations to evaluate and validate the method.
Details are presented in the following section.

\subsection{Comparison with the method presented by \cite{yildirimoglu_experienced_2013}}

As already mentioned, our method share several common features with the work of \cite{yildirimoglu_experienced_2013}.
Consequently, this section proposes a short comparison of the two approaches.

To summarize, the method of \cite{yildirimoglu_experienced_2013} aims to identify groups of days presenting similar traffic conditions to build stochastic congestion maps, identifying the probability that a section of the test case is congested. 
According to the newly available information, the stochastic congestion maps are used to revise the state of a priori knowledge.
Their work is organized as follows:
\begin{itemize}
    \item Historical data:
    \begin{itemize}
        \item Groups of similar days are identified with a PCAxGMM combination (speed is the observation vector);
        \item Stochastic congestion maps are produced for each group to highlight blocks of recurrent congestion;
    \end{itemize}
    \item Real-time: the goal is to identify the closest situation in the historical data to determine which bottlenecks are active. Then, potential active bottlenecks are integrated in the processing of real-time measurements to produce experienced travel times.
    \begin{itemize}
        \item Identification of the active blocks based on the comparison of the real-time congestion map with stochastic congestion maps;
        \item Travel times are produced by combining the real-time measurements of the speed at the different detectors and a correction based on the identification of potential bottleneck.
    \end{itemize}
\end{itemize}

In comparison, our method aims to identify groups of days presenting similar traffic conditions to determine a single representative day of each group. 
The measurements of these representative days are then used in real-time to predict future traffic conditions. 
The newly available information is only used to determine the closest representative days and is directly processed to product prevision of the travel times. 
The method can be resumed as follows:
\begin{itemize}
    \item Historical data:
    \begin{itemize}
        \item Groups of similar days are identified with a PCAxGMM / k-means combination (speed in the observation vector);
        \item Identification of the most representative day of each group by consensual learning and congestion maps;
    \end{itemize}
    \item Real-time: the goal is to identify the closest consensual day and use past observations to predict future conditions.
    \begin{itemize}
        \item Identification of the closest consensual day by the comparison of real-time congestion map with the consensual congestion maps;
        \item Travel times are directly those observe for the identified consensual day.
    \end{itemize}
\end{itemize}

\section{Clustering of days with similar traffic conditions into the historical dataset}

\subsection{Determining the optimal number of clusters}

For both approaches, the number of groups $K$ has to be fixed exogenously, i.e., before performing the clustering of the dataset.
Even if there is no definitive answer, several methods exist to determine the optimal number of clusters.
These methods are either based on a criterion minimization/maximization (such as the elbow or averaged Silhouette methods \citep{ROUSSEEUW198753}) or on a statistical test (such as gap statistic method, Bayesian information criterion or Akaike information criterion).  
However, these criteria are frequently not consistent between them, and it was the case for our study.
Consequently, we have decided to tailor our metrics to determine the optimal number of clusters $K$ that we need to predict the evolution of congestion and travel times.

The prerequisite is to determine clusters that gather a sufficient number of days (for example, $5\%$ of the test dataset) with similar traffic conditions (intra-cluster homogeneity) and that are different enough from one to the other to justify different groups (inter-cluster dissimilarity). 
To this end, three metrics have been developed. 
The intra-cluster homogeneity is appraised by calculating the average of the Rand index inside a cluster:
\begin{equation}
    \frac{1}{|C_k|} \sum_{k \in K} \frac{1}{|C_k|-1} \sum_{\forall (p,q) \in C_{k}, p \neq q} R_M(p,q),    
\end{equation}
where $K$ is the number of clusters, $C_k$ is the cluster $k$, $d_k$ is the consensual day of cluster $C_k$, $p$ and $q$ are days belonging to $C_k$, and $|C_k|$ is the cardinal of cluster $C_k$.
The inter-cluster dissimilarity is evaluated by determining the average of the Rand index between each pairs of consensual days:
\begin{equation}
    \frac{1}{\binom{n}{2}} \sum_{\forall (p,p') in D_{k}} R_M(p,p').
\end{equation}
Finally, we also compared the number of clusters $n_{min}$ gathering more than $5$ days to the targeted number of cluster $K$.

In order to perform this analysis, the clustering process is iterated $20$ times to ensure the generality of the results.
Then, the averages of the three metrics presented above are calculated.

\begin{figure}[H]
    \centering
        \includegraphics[width=\textwidth]{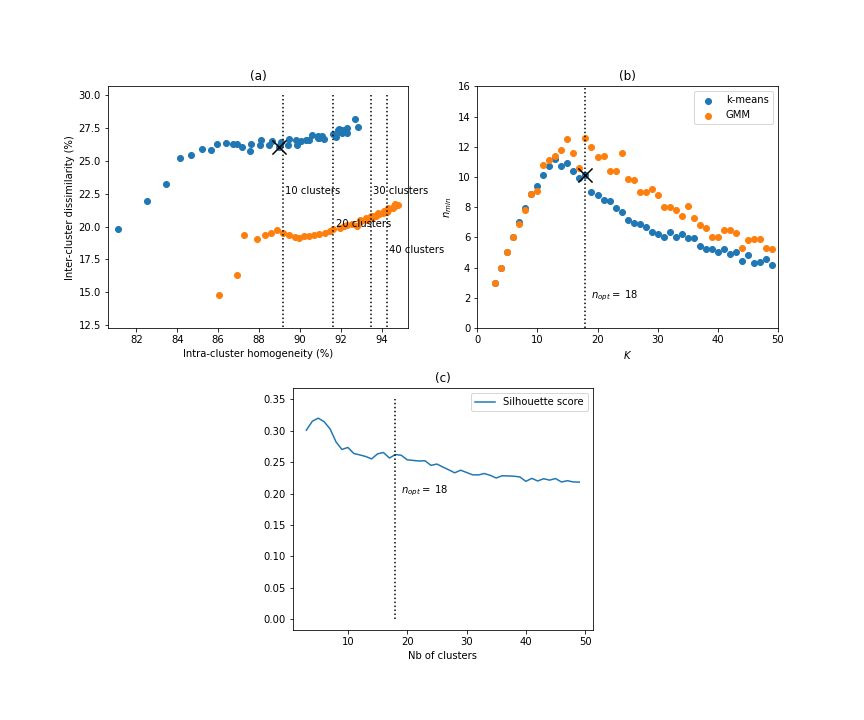}
        \caption{(a) Comparison of intra-cluster homogeneity and  inter-cluster dissimilarity, (b) Number of large clusters vs $K$ for the k-means and the GMM methods, and (c) Evolution of the Silhouette score with the number of clusters}.
        \label{fig:fig2b}
\end{figure}

Figure \ref{fig:fig2b}a compares the intra-cluster homogeneity with inter-cluster dissimilarity in regards to various values of $K$ for both approaches. 
It reveals that the inter-cluster dissimilarity is stable for $K > 10$.
Simultaneously, the intra-cluster homogeneity continues to increase but a very low rate.
In the meantime, Figure \ref{fig:fig2b}b depicts $n_{min}$ the number of clusters gathering more than $5$ days in function of $K$. 
We also propose the evolution of the Silhouette score with the number of clusters (see Figure \ref{fig:fig2b}c).
It clearly shows that for $K$ bigger than $30$, both methods are unable to identify large clusters.
It is even worse for high values of $K$.
Consequently, it appears that $n_{opt} = 18$ constitutes a good balance to gather the data of our test site (black cross in Figure \ref{fig:fig2b}a, b, and c).
Especially, we can almost observe a local optimum for the average silhouette width at between $15$ and $20$ clusters.
$n_{opt} = 18$ ensures a sufficient number of clusters and a sufficient number of days per cluster (more than $13$ days on average) with an acceptable intra-cluster homogeneity and inter-cluster dissimilarity.
It also means that clusters can be modified or removed retrospectively without changing the metrics significantly.

In addition to these averaged metrics, the stability of the clustering results is also studied. 
To this end, the Rand index between successive partitions is calculated. 
Consequently, if $(C_{n-1})$ is the $n-1$ partition and $(C_n)$ is the $n$ partition, the Rand index can be expressed as follows:
\begin{equation}
    R = \frac{a+b}{a+b+c+d}
\end{equation}
where:
\begin{itemize}
    \item a is the number of pairs of days that are in the same clusters in $(C_{n-1})$ and $(C_n)$;
    \item b is the number of pairs of days that are different clusters in $(C_{n-1})$ and $(C_n)$;
    \item c is the number of pairs of days that are in the same clusters in $(C_{n-1})$ and in different clusters in $(C_n)$;
    \item d is the number of pairs of days that are in different clusters in $(C_{n-1})$ and in the same clusters $(C_n)$.
\end{itemize}

\begin{figure}[H]
    \centering
        \includegraphics[width=\textwidth]{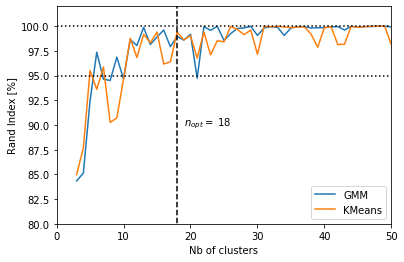}
        \caption{Evolution of the Rand Index between successive partitions}.
        \label{fig:fig2c}
\end{figure}

Figure \ref{fig:fig2c} shows the evolution of the Rand index with the number of clusters for the two clustering methods.
From 10 to 50 clusters, less than $5\%$ of days change group when increasing the number of clusters.
It appears that the results are reasonably stable and confirm that $n_{opt} = 18$ is a satisfying trade-off between the different metrics.

\subsection{Post-clustering analysis}

\begin{figure}[H]
    \centering
        (a) k-means method
        \includegraphics[width=\textwidth]{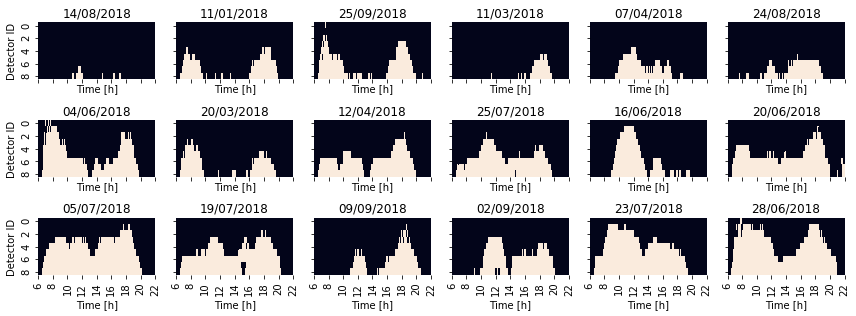}
        (b) GMM method
        \includegraphics[width=\textwidth]{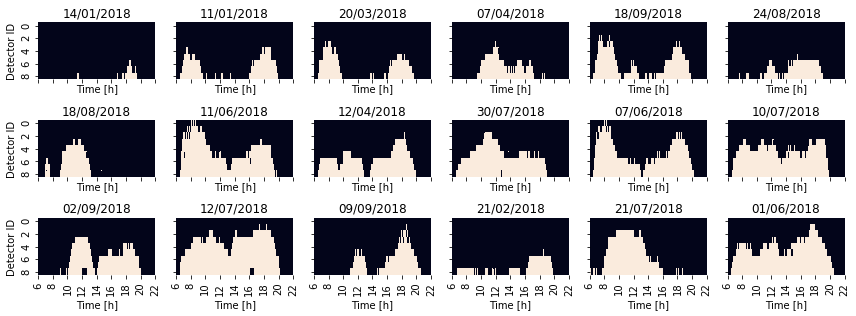}
        \caption{Congestion map of consensual days $d_k$ identified with (a) k-means method and (b) GMM method}.
        \label{fig:jour_con}
\end{figure}

First of all, Figure \ref{fig:jour_con}a (b, respectively) shows the congestion maps of the consensual days $D_k$ that have been obtained for the k-means method (GMM, respectively).
Interestingly, the patterns are quite different from one day to the other, for both methods. 
Note that $d_k$ have been sorted by decreasing size of $C_k$, i.e. $C_i$ gathers more days than $C_{i+1}$.
Notice that the consensual days of the k-means are denoted $d_i$ whereas those of GMM method are denoted $d'_i$.
Congestion maps of $d_1$ (\ref{fig:jour_con}a) for k-means clustering and of $d'_1$ for GMM method (\ref{fig:jour_con}b) clearly correspond to the cluster of free-flow days.
Situations with morning and evening peak hours can be identified (see, for example, $d_3$ and $d'_2$) or with only a morning peak hour (see, for example, $d_{4}$ and $d'_{15}$).
It is also important to notice that the two methods do not lead exactly to the same consensual days (and the same clusters).
In this case, they only have $7$ consensual days in common.

\begin{figure}[H]
    \centering
        \includegraphics[width=\textwidth]{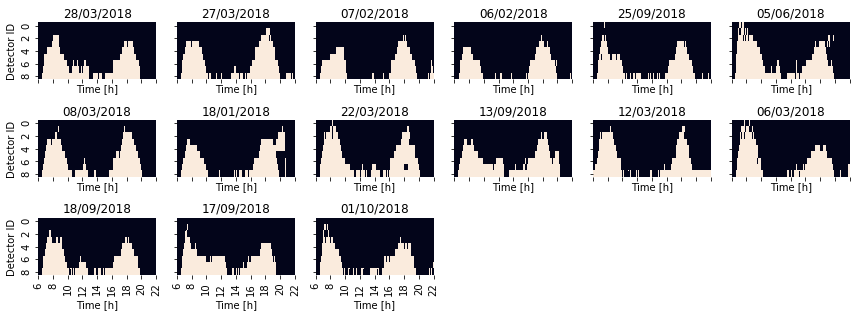}
        \caption{Congestion maps cluster $C_3$ obtained with the k-means method}.
        \label{fig:C3}
\end{figure}

Figure \ref{fig:C3} shows the different congestion maps of the days gathered into cluster $C_3$.
Visually, the pattern is similar with always morning and evening peak hours.

Interpreting the results of the clustering methods is always a difficult task because a relevant balance must be found between a trivial classification and useless and complex justifications of the hypothetical links between observations grouped into the same cluster.
In our case, the number of attributes available to analyze and explain the configuration of the computed clusters is very limited.
Consequently, the post-clustering analysis is mainly focused on the weekday and the month of the observations that are gathered into the same group.

\begin{figure}[H]
    \centering
        \includegraphics[width=\textwidth]{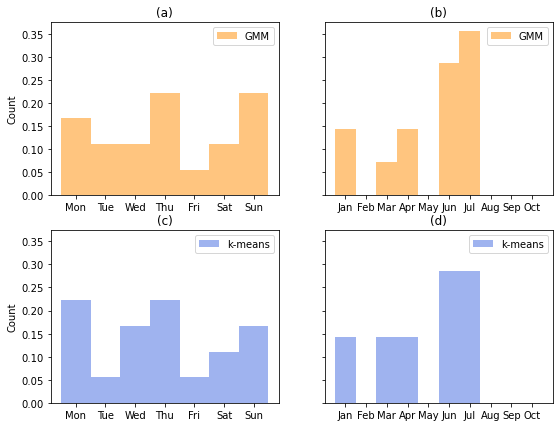}
        \caption{Distribution of weekdays and months of identified consensual days for the (a-c) GMM method and (b-d) k-means method}.
        \label{fig:JC}
\end{figure}

Figure \ref{fig:JC} shows the distributions of both variables.
It turns out that distributions among days (Figure \ref{fig:JC}a and c) and months (Figure \ref{fig:JC}b and d) are very similar.
Note that several iterations have been performed for each method and that the histograms have been normalized.

\begin{figure}[H]
    \centering
        \begin{tabular}{ccc}
            \includegraphics[width=0.3\textwidth]{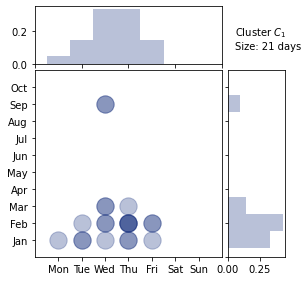} & \includegraphics[width=0.3\textwidth]{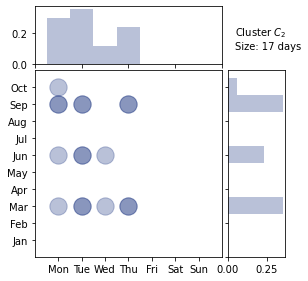} & 
            \includegraphics[width=0.3\textwidth]{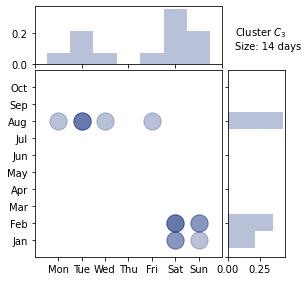} \\
            \includegraphics[width=0.3\textwidth]{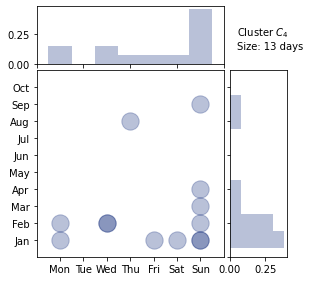} & \includegraphics[width=0.3\textwidth]{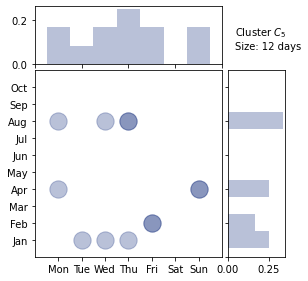} & 
            \includegraphics[width=0.3\textwidth]{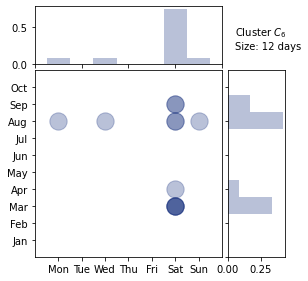} 
        \end{tabular}
    \caption{Distribution of weekdays and months of the days of the $6$ largest clusters identified with k-means method}.
    \label{fig:distrib_kmeans}
\end{figure}

To continue the post-clustering analysis, distributions of weekdays and months of the days gathered into the same clusters are now analyzed. 
Because the results are very similar for both approaches, the study is now centered on the k-means method.
Let us focus on the six largest clusters of the partition, see Figure \ref{fig:distrib_kmeans} that shows the normalized distributions of the weekdays and the months.
Cluster $C_1$ mainly gathers Saturdays and Sundays, and the month of August, that corresponds to the summer holidays in France.
Without any surprise, the visual inspection of the congestion maps reveals that this group is composed of free-flow situations, see the congestion map of $d_1$ in Figure \ref{fig:jour_con}.
Cluster $C_2$ encompasses the weekdays, mainly during winter.
Furthermore, Cluster $C_3$ is very interesting because it almost only holds Saturdays of holidays: February/March for winter holidays and August for summer holidays.
Cluster $C_5$ gathers the beginning of the week (Mondays and Tuesdays), whereas Cluster $C_6$ the end of the week.
The comparison of the congestion maps of the different days, and $d_5$ and $d_6$ reveals different patterns: the midday hours are more congested for $C_5$.
Finally, cluster $C_4$ groups days with only evening congestion; it may explain the majority of Sundays in the set.

\begin{figure}[H]
    \centering
        \includegraphics[width=\textwidth]{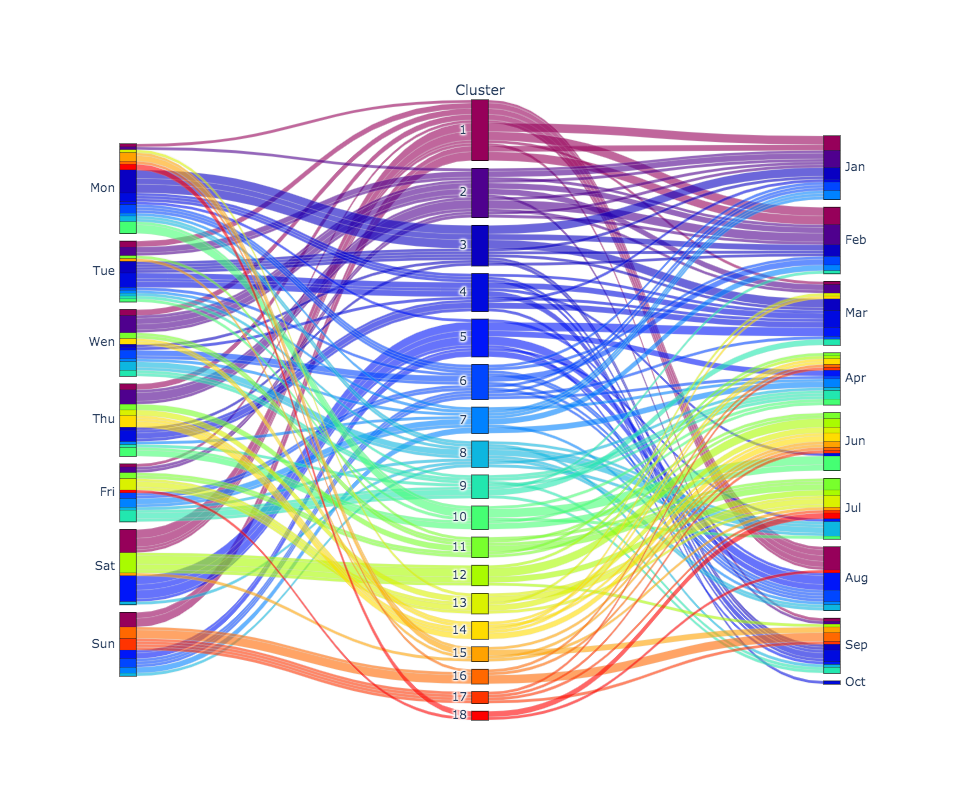}
        \caption{Alluvial diagram of contribution of days and months into each cluster}.
        \label{fig:sankey}
\end{figure}

Because analyzing the results cluster by cluster is a tedious task, a convenient approach is to use an alluvial diagram.
Figure \ref{fig:sankey} shows the contribution of each weekday and month into the different clusters.
The careful analysis of the alluvial diagram completes the previous analysis.
For example, it appears that Sundays are only present in 5 clusters, such as the Saturdays.
On the contrary, each weekday can be found in $10$ different clusters.
A similar analysis can be performed with the contribution of the months to each cluster.

These results are very encouraging: the GMM and the k-means methods capture physically similar situations.
Because the two approaches are fully explainable and rely on understandable metrics, all the identified clusters can be easily justified and interpreted.

\section{Results}
The global prediction method is now tested for our case study.
We mainly focus on two metrics: the predicted congestion maps and the travel times series.
As already explained, the data has been divided into a training set ($75\%$ of the initial data) and a validation set (the remaining $25 \%$).
Moreover, we decided to only focus on the k-means method for the sake of brevity in the presentations of the results.

\subsection{Comparison with existing method}

We first decide to compare our method with existing approaches to identify the optimal domain of application.
This task is tough for many reasons. 
Notably, it necessitates selecting an existing method that is accurate enough, and which does not require tremendous work to be developed and calibrated. 
Moreover, existing methods have different objectives and domains of applications. Consequently, we decided to compare our approach to a naive instantaneous method and a historical average method. 
Besides, the comparison with two other versions of our method is performed to convince about the predictive power of our approach.

Therefore, we define the following methods:
\begin{itemize}
    \item $M0$ is a naive approach consisting in shifting the observation made at time $t$ to a horizon $\delta t$. Consequently, this method will be almost perfect for short $\delta t$;
    \item $M1$ is a historical average method consists in calculating average values of the historical data for each day of the week and each time period. 
    \item $M2$ is the original method proposed in the paper based (Figure \ref{fig:fig0}) on the clustering of the historical data and the identification of consensual days;
    \item $M3$ is the same as $M2$ except that averaged congestion maps are calculated for each group instead of identifying a consensual day;
    \item $M4$ follows the same process as $M2$ except that historical data are not clustered. 
    Each day of the historical dataset can be used as a prediction.
\end{itemize}

To compare these approaches, we use the following metrics:
\begin{itemize}
    \item congestion prediction: Rand indexes, i.e. accuracy, and F1-score between predicted congestion maps and ground truth;
    \item Travel times prediction: RMSE between prevision and observation.
\end{itemize}

The test set is composed of the 56 days that have not been used to perform the clustering (167 days in the historical data set).

\begin{figure}[H]
    \centering
        \includegraphics[width=1\textwidth]{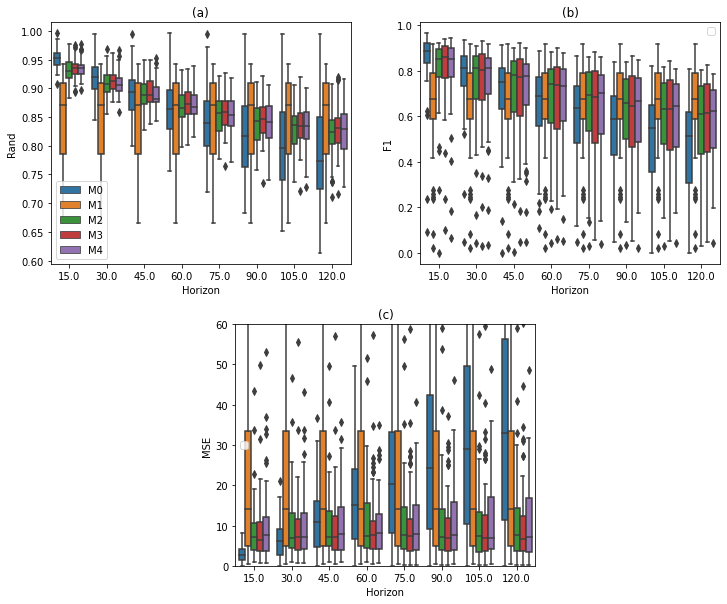} 
    \caption{Evolution of the boxplots of (a) Rand index, (b) F1-score, (c) MSE according to the horizon (in minutes) between predictions and observations}.
    \label{fig:comp1}
\end{figure}

Figure \ref{fig:comp1}a (respectively b) shows the evolution of the boxplots of the rand index (respectively F1-score) between predicted and observed congestion maps for the four different methods according to the horizon predictions. 
It turns out that methods $M2$, $M3$, and $M4$ outperform the instantaneous method $M0$ for horizons bigger than $30$ minutes.
In the same time, these methods are better than the historical approach $M1$ for horizons smaller than $1$ hour.
Figure \ref{fig:comp1}c depicts similar results for the MSE between predicted and observed travel times, except that that $M1$ is almost the worst method between the five tested. 
The comparison shed light on the optimal domain of application of our method. 
Consequently, a horizon of $\delta t = 60$ minutes is used in the remaining of the paper.

It is also important to notice that methods $M1$, $M2$, and $M3$ have similar performance concerning the congestion propagation prediction. 
This is not surprising because they are three variants of the same methods. 
Besides, method $M2$ and $M1$ are slightly better than $M3$ for the travel times.
However, we prefer to retain method $M1$ because the clustering makes it more understandable (prediction can be related to qualitative transportation scenario), and the use of a consensual day produces more realistic travel times when focusing on the time series of the predicted travel times.

\subsection{Congestion propagation}
The ability of the proposed method to anticipate congestion propagation is now appraised by comparing the predicted congestion maps with the real observations.
Remember that the prediction method has three parameters: the duration $\Delta t$ that is used to make the prediction, the horizon $\delta t$ for which the prediction is made, and the congested speed threshold (fixed here to $40$ km/h).

\begin{figure}[H]
    \centering
        (a) \hspace{2cm} (b) \hspace{2.5cm} (c) \hspace{2cm} (d)
        \includegraphics[width=\textwidth]{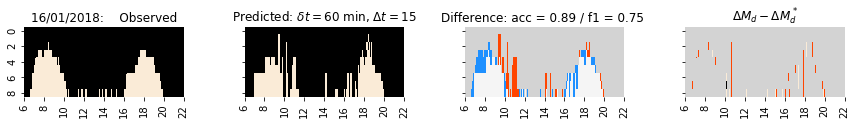}
        \includegraphics[width=\textwidth]{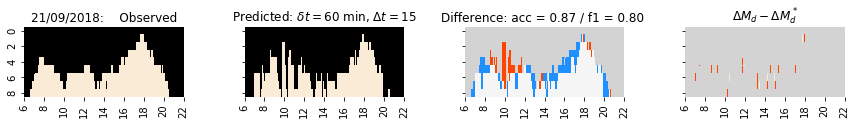}
        \includegraphics[width=\textwidth]{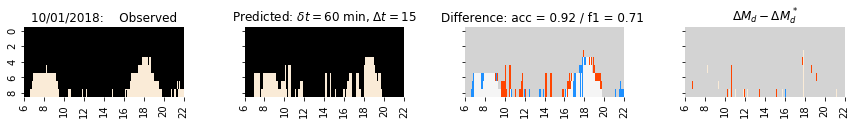}
        \includegraphics[width=\textwidth]{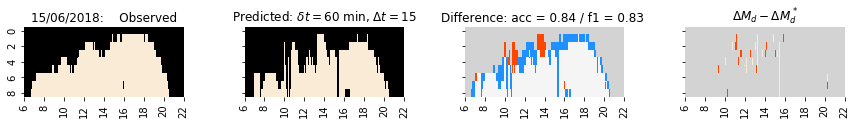}
        \includegraphics[width=\textwidth]{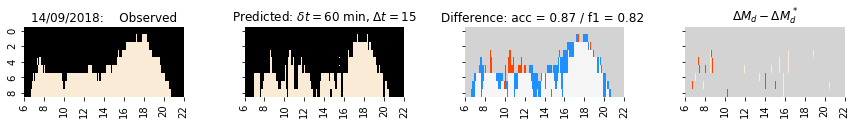}
        \includegraphics[width=\textwidth]{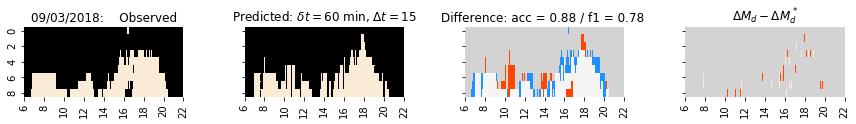}
        \caption{Comparison of (a) observed and (b)predicted congestion maps (black is for free-flow - FF conditions, beige for congestion - C Conditions), (c) difference between them (blue is for C / FF and red for FF /C , white and gray for correct predictions) and (d) operational metrics of prediction accuracy (red is for +1, blue -1, black - 2, beige +2 and gray for correct predictions)}
        \label{fig:fig5}
\end{figure}

Figure \ref{fig:fig5} shows the observed maps, the predicted congestion maps (for $\Delta t = 15$ min and $\delta t = 60$ min), and the difference between these maps for $6$ randomly selected days of the testing sample.
Note that the predicted congestion maps are the result from the repeated previsions at every time step (one minute in our case).
The visual inspection reveals that the differences are very low between prediction and ground truth, and this for significantly different shapes of congestion propagation (Figure \ref{fig:fig5}c, the blue color is when a congestion propagation is predicted, but the observation is free-flow, red for the opposite).
This qualitative analysis is very encouraging.

To go further, we decided to calculate the accuracy of the prevision. 
The values are particularly good since they are very close to $100\%$, see titles of Figure \ref{fig:fig5}.
This score might be imputed to the free-flow situations that are the most frequent and easy to predict in the congestion maps.
We also calculate the F1-score of the prevision, that are all satisfying even if this metrics is more sensitive to the wrong prediction.
However, it should be noted that the F1-score is calculated with the congested traffic conditions only.
It may introduces a bias when comparing days with significant different volume of congestion.
However, the forthcoming travel times evolution analysis will confirm that the results of the method are encouraging.

Keeping in mind that the method proposed here is tailored to be practice-ready and answer to operational needs, specific metrics are developed to evaluate the prediction accuracy.
The confidence that can be allowed by an operator to a decision support system, such as a traffic evolution predictor, is often based on simpler indicators than those used in a research paper. 
A potential criterion is the decision support system's ability to predict in the same direction what can be observed in real-time by the operator.
For example, if an increase of the congestion is predicted for a horizon $\delta t$, will the operator really observe the congestion propagation in his.her monitoring system a time $\delta t$ later?

Consequently, only predictions of variations are now evaluated. 
To this end, the difference $\Delta M_{d} = M_{d}(i+1,j)-M_{d}(i,j)$ can be calculated for all $i \in [1,n-1]$.
This variable is equal to $0$ if queue length remains stable, $1$ if the congestion propagates upstream, and $-1$ if the queue length reduces.
Then, $\Delta M_{d}$ can be compared to the prediction $\Delta M^*_{d}$.
It leads to the following table of the values of $\Delta M_{d}-M^*_{d}$.

\begin{table}[H]
\centering
\begin{tabular}{|c|c|c|c|c|}
\hline
\cellcolor[HTML]{EFEFEF} minus    & \multicolumn{4}{c|}{\textbf{$\Delta M*_{d}(i,j)$}}                                                                                                                     \\ \hline
                             & \cellcolor[HTML]{C0C0C0}\textbf{-}  & \cellcolor[HTML]{C0C0C0}\textbf{0} & \cellcolor[HTML]{C0C0C0}\textbf{1} & \cellcolor[HTML]{C0C0C0}\textbf{-1} \\ \cline{2-5} 
                             & \cellcolor[HTML]{C0C0C0}\textbf{0}  & 0                                  & -1                                 & 1                                   \\ \cline{2-5} 
                             & \cellcolor[HTML]{C0C0C0}\textbf{1}  & 1                                  & 0                                  & 2                                   \\ \cline{2-5} 
\multirow{-4}{*}{\textbf{$\Delta M_{d}(i,j)$}} & \cellcolor[HTML]{C0C0C0}\textbf{-1} & -1                                 & -2                                 & 0                                   \\ \hline
\end{tabular}
\caption{Comparison of the evolution of predicted and observed congestion lengths based on the congestion maps}
\label{tab:tab1}
\end{table}

Figure \ref{fig:fig5}d shows the colormap of $\Delta M_{d}- \Delta M^*_{d}$ for the $6$ random selected day.
It clearly reveals that only a few errors about the evolution direction are made by the method.
Especially if we compare to the rough differences of the prediction and the observation (Figure \ref{fig:fig5}d), we do not accumulate the errors.
In addition to this qualitative analysis, a global indicator can be computed: the ratio $\rho$ of accurate prediction over the total number of observations.

\begin{equation}
    \rho(d) = \frac{card \Big ( \{ \Delta M_{d}- \Delta M^*_{d}= 0 \} \Big ) }{n-1}
\end{equation}

Distribution of $\rho$ for the whole testing set is not shown because it is completely centered on the mean value ($\bar{\rho} = 98.2 \%$) with a minimal standard deviation ($0.4 \%$). 
Therefore, the accuracy in the prediction of the evolution of the traffic conditions is excellent.
It is very encouraging for the potential integration of the method into an operational decision support system. 
This will be the case for our test study.

\subsection{Travel times estimation}

As already mentioned, the proposed method can predict travel times by using the observed speeds of the consensual days.
The main benefit of this approach is to produce realistic travel times because predictions come from past observations.

\begin{figure}[H]
    \centering
        \includegraphics[width=\textwidth]{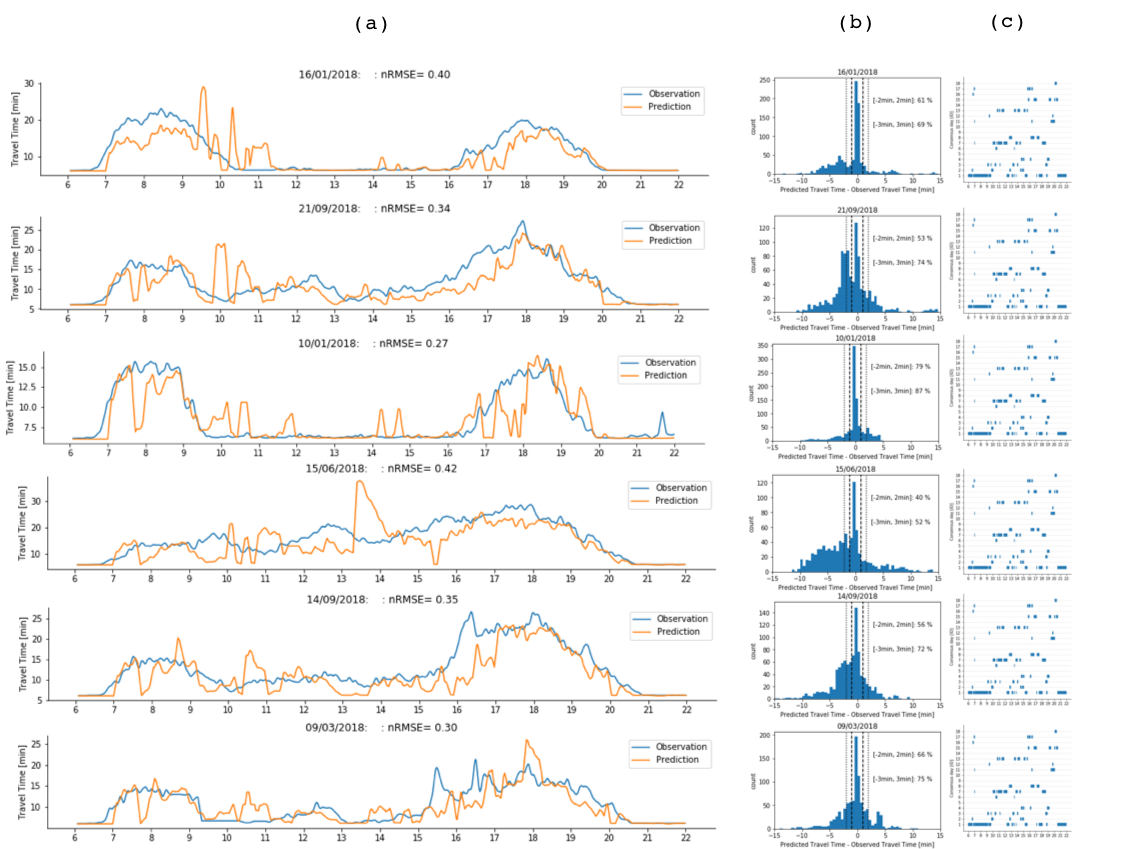}
        \caption{(a) Comparison of observed (in blue) and predicted travel times (in orange), (b) distribution of the absolute errors, (c) evolution of optimal consensual days along the prediction}
        \label{fig:fig6}
\end{figure}

Figure \ref{fig:fig6}a shows the travel times series for the $6$ previously studied days. 
The orange curves correspond to the prediction, whereas the blue ones are the observations.
Note that a horizon $\delta t = 60$ min and a learning duration of $\Delta t = 15$ min have been used.

Visually, the results are very good, and the trends of traffic evolution are well predicted.
The method can accurately capture the variations of travel times. 
Besides, we have calculated the normalized RMSE between observations and predictions, and the criterions appear to be satisfying.
To complete the analysis, Figure \ref{fig:fig6}b shows the distributions of the errors between predictions and observations.
The proportion of the predictions that have a precision below $2$ minute stands from $40\%$ to $79\%$.
If we focus on the $3$ minutes window, this value increases to $68 \%$ to $87 \%$.
Finally, Figure \ref{fig:fig6}c highlights the variation of the optimal consensual day determined by the prevision method.
It shows out that a large subset of the consensual days is used to predict travel times throughout the day.

\section{Conclusion}

Prediction of congestion propagation and travel times evolution is still a topic actively studied in the literature.
This paper tries to make its contribution by proposing a simple method, which has the main benefit of being fully explainable compared to recent approaches based on machine learning or artificial intelligence methods.

The key component is the concept of \textit{congestion map}, which is a binary observation metric of traffic states on a highway.
Using this proxy reinforces the importance of focusing on traffic dynamics rather than on the numeric values of observation variables. 
Thus, the historical dataset of a French $10$ km long freeway is classified into groups of days presenting similar traffic states.
The second innovation of the proposed method is to identify a \textit{consensual day} for each cluster. 
According to a distance based on the Rand index, the objective is to determine the day that is the most representative of the cluster.
It makes it possible to find almost the expected traffic situations of this highway: morning and evening peak hours, only morning/evening peak hours, all day long congestion, free-flow day, holiday traffic, etc. 

Once those consensual days have been determined, the method can be applied in real-time to predict congestion propagation and travel times evolution.
According to a real-time learning period, observations of a new day are compared to the consensual days' congestion maps. 
The closest one is identified, and the congestion map and speeds that have been observed for this specific day are used to predict the behavior of the new day for a given horizon.
This very simple method gives encouraging results for both congestion and travel times evolution.
Especially, the comparison with naive methods (instantaneous and historical average) reveals that the proposed model is useful for longer prediction horizons.

Various future directions can be pursued.
The methodology can be improved by identifying the best duration to compare congestion maps, the accurate prediction horizon, and the congested speed threshold.
To perform the study, we naively tested different values, but a sensitivity analysis could be conducted. 
A second improvement could be to use congestion maps that are no more dependent on the time of the day.
The idea is to focus only on the shape of the maps, i.e., shockwaves profiles. 
These claims still need to be researched and validated.

\newpage
\section*{Acknowledgments}

This research work was carried out as part of the “Lyon Covoiturage Experimentation – LCE” project at the Technological Research Institute SystemX, and was supported with public funding within the scope of the French Program “Investissements d’Avenir”.

The authors thank the different members of the IRT-Sytem X for the fruitful discussion about the data clustering and the Metropole de Lyon, which provide the data for this study. 
Plenty of other interesting datasets can be found on: https://data.grandlyon.com/

This paper was mainly written during the long night of the Covid-19 lockdown.

\newpage
\section*{List of notations}
\begin{table}[H]
\setcellgapes{2.5pt}
\renewcommand*{\arraystretch}{1}
\makegapedcells
\begin{tabular}{ | c  l  l | }
  \hhline{|===|}
  Variable			& Description 							&\\
  \hhline{|===|}
  $t$ 			& Time of the day					& \\
  $l$ & Loop detector index &\\
  $\Delta x_l$ & Length of section $l$ & \\
  $x_l(t)$ & Binary variable & \\
  $v_l(t)$ & Speed recorded at detector $l$ at time $t$ & \\
  $v_{cong}$ & Congestion speed threshold & \\
  $M_d$ & Boolean matrix of congestion map of day $d$ such as $M_d(l,t) = x_l(t)$ & \\
  $R_M(d,p)$ & Rand index between congestion maps of day $d$ and $p$ & \\
  $C_k$ & Cluster $k$ & \\
  $K$ & Number of clusters & \\
  $d_k$ & consensual day of cluster $C_k$ & \\
  $D_k$ & Set of consensual days & \\
  $\Delta t$ & Learning period & \\
  $\delta t$ & Horizon of prediction & \\
  $d^*_p(t)$ & Optimal consensual day for a new observation day $p$ at time $t$ & \\
  \hhline{|===|}
\end{tabular}
\label{notations}
\nomakegapedcells
\end{table} 
\newpage
\bibliography{Heart.bib} 

\end{document}